\documentclass[journal=nalefd, manuscript=letter, layout=twocolumn]{achemso}

\usepackage[version=3]{mhchem} 
\usepackage{graphicx}
\usepackage{dcolumn}
\usepackage{bm}
\usepackage{xr}
\usepackage{color}
\usepackage{amssymb,amsmath}

\author{M. Fernando Gonzalez-Zalba}
\email{mg507@cam.ac.uk}
\affiliation{Hitachi Cambridge Laboratory, Cambridge, UK}
\author{Sergey N. Shevchenko}
\affiliation{B.Verkin Institute for Low Temperature Physics and Engineering, Kharkov, Ukraine}
\alsoaffiliation{V. Karazin Kharkov National University, Kharkov, Ukraine}
\alsoaffiliation{Center for Emergent Matter Science, RIKEN, Wako-shi, Saitama 351-0198, Japan}
\author{Sylvain Barraud}
\affiliation{CEA/LETI-MINATEC, CEA-Grenoble, Grenoble, France}
\author{J. Robert Johansson}
\affiliation{Center for Emergent Matter Science, RIKEN, Wako-shi, Saitama 351-0198, Japan}
\author{Andrew J. Ferguson}
\affiliation{Cavendish Laboratory, University of Cambridge, UK}
\author{Franco Nori}
\affiliation{Center for Emergent Matter Science, RIKEN, Wako-shi, Saitama 351-0198, Japan}
\alsoaffiliation{Department of Physics, The University of Michigan, Ann Arbor, Michigan, USA}
\author{Andreas C. Betz}
\affiliation{Hitachi Cambridge Laboratory, Cambridge, UK}

\date{\today}

\title{Gate-sensing coherent charge oscillations \\ in a silicon field-effect transistor}


\AbstractOn
\begin{document}

\begin{abstract}
Quantum mechanical effects induced by the miniaturization of complementary metal-oxide-semiconductor (CMOS) technology hamper the performance and scalability prospects of field-effect transistors. However, those quantum effects, such as tunnelling and coherence, can be harnessed to use existing CMOS technology for quantum information processing. Here, we report the observation of coherent charge oscillations in a double quantum dot formed in a silicon nanowire transistor detected via its dispersive interaction with a radio-frequency resonant circuit coupled via the gate. Differential capacitance changes at the inter-dot charge transitions allow us to monitor the state of the system in the strong-driving regime where we observe the emergence of Landau-Zener-St{\"u}ckelberg-Majorana interference on the phase response of the resonator. A theoretical analysis of the dispersive signal demonstrates that quantum and tunnelling capacitance changes must be included to describe the qubit-resonator interaction. Furthermore, a Fourier analysis of the interference pattern reveals a charge coherence time, $T_2\approx 100$~ps. Our results demonstrate charge coherent control and readout in a simple silicon transistor and open up the possibility to implement charge and spin qubits in existing CMOS technology.
\end{abstract}

\newpage

Quantum computation promises to be exponentially more efficient than classical computers in solving a particular set of problems~\cite{Feynman1982,Shor1997,Grover1997}. However, implementing the underlying quantum algorithms requires a scalable hardware that would allow making multi-qubit structures possible. Silicon quantum-dot-based qubits are promising candidates for such quantum hardware due to their tunability, flexible coupling geometries and long coherence times~\cite{Veldhorst2014,Kim2014,Veldhorst2015}. Furthermore, using silicon one can exploit the advances of CMOS technology and benefit from an industrial platform dedicated to building complex scalable circuits.

A first step towards quantum computation with CMOS quantum dots would be demonstrating that time-dependent coherent phenomena can be harnessed in a scalable CMOS device. One approach to test the coherent nature of a system is Landau-Zener-St{\"u}ckelberg-Majorana (LZSM) interferometry~\cite{Nakamura2012,Shevchenko2010}, in which a coupled two-level system is strongly driven through its anti-crossing. This approach has been successfully applied for coherent quantum control of superconducting qubits~\cite{Oliver2005,Berns2006,Sillanpaa2006,Wilson2007,Berns2008,LaHaye2009,deGraaf2013,Quintana2013}, semiconductor quantum dots~\cite{Petta2010,Stehlik2012,Ribeiro2013,Cao2013} and donors in silicon~\cite{Dupont-Ferrier2013}.

Additionally, interfacing quantum systems with high-frequency electrical resonators promises compact high-sensitivity quantum-state read-out and long distance transfer of information~\cite{Wallraff2004,Petersson2012,deLange2015,Larsen2015}, ideal characteristics for a prospective scalable architecture. In these systems, the dispersive shift on the resonator response due to the qubit's state-dependent quantum or tunnelling capacitance ~\cite{Sillanpaa2005,Duty2005, Ashoori1992, Ciccarelli2012,Colless2013,Gonzalez-Zalba2015} is exploited for read-out. However, in the strong-driving regime these two different dispersive contributions can co-exist~\cite{Sillanpaa2006} and it becomes important to understand the nature of the qubit-resonator interaction and the different contributions to the dispersive response.

Here, we demonstrate coherent control and read-out of the charge state of a double quantum dot (DQD) in a CMOS transistor. We perform dispersive charge detection \textit{in-situ} by coupling the gate of the transistor to a MHz resonator and monitoring changes in the differential capacitance at the inter-dot charge transitions. We show coherent manipulation of the charge state in the strong-driving regime, where we observe LZSM interference on the charge occupation probabilities of the DQD. We find that the DQD-resonator interaction is accurately described by a combination of quantum capacitance changes, due to the non-zero energy-band curvature, and tunnelling capacitance variations, since the quantum state probability redistribution happens at a rate much faster than the probing frequency of the resonator. Finally, we obtain the charge coherence time by analyzing the interference signal in Fourier space. Overall, our work demonstrates charge coherent manipulation and read-out in a CMOS transistor, paving the way towards CMOS-based quantum computing.  

\begin{figure}[htbp]
	\includegraphics{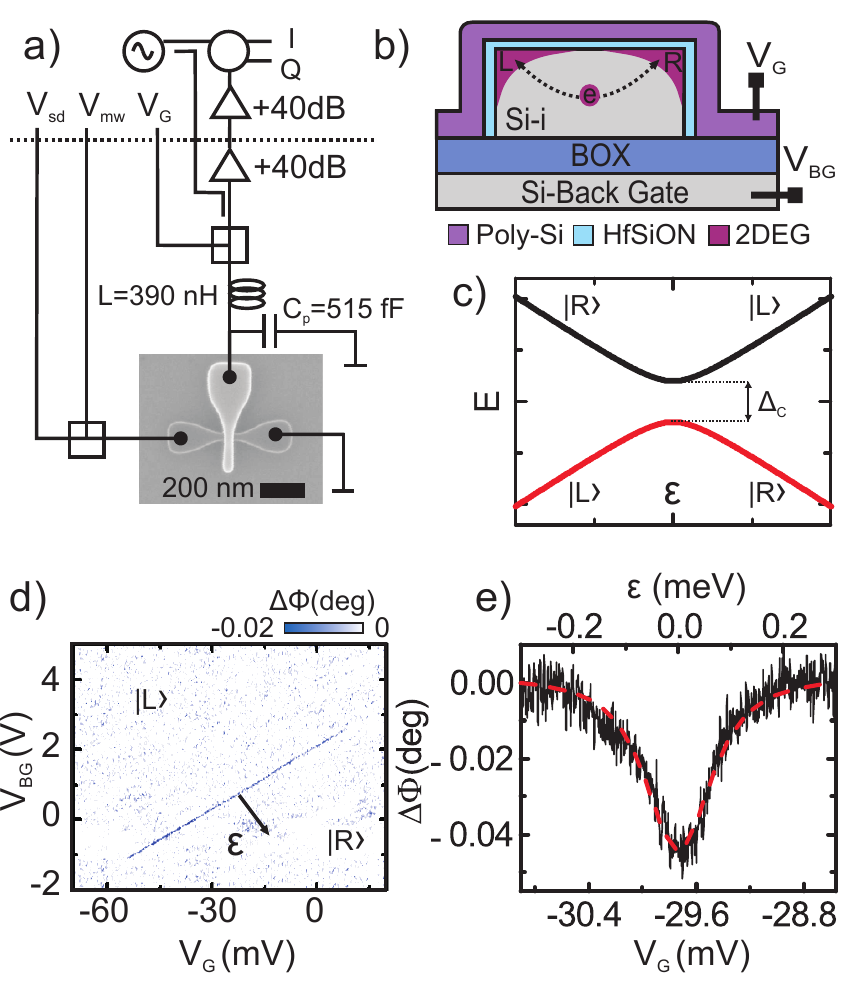}
	\caption{{\bfseries Device characterization and measurement of inter-dot quantum capacitance}.(a) Scanning electron microscope image of a similar transistor (channel width $w=80$~nm and gate lentgh $l=50$~nm) connected to a reflectometry set up via the gate. DC and MW voltages are delivered to the source via a K250 bias-tee. The MW line is attenuated -23~dB at 1~K and ~-3~dB at 45~mK. (b) Schematic cross-section of the transistor perpendicular to the transport direction. Quantum dots L and R form at the top-most edges of the transistor due to the corner effect. (c) Schematic energy diagram of a single-electron shared among two tunnel coupled quantum dots as a function of the energy detuning $\varepsilon$. The charge state configuration of the ground state (red curve) and excited state (black curve) are indicated as $\left|L\right\rangle$, $\left|R\right\rangle$. (d) $V_\text{G}$-$V_\text{BG}$ stability map of the phase response of the resonator where an inter-dot charge transition is observed. The direction of increasing detuning is marked by a black arrow. (e) Phase response (black solid curve) and fit (red dashed curve) of the inter-dot transition as a function of $V_\text{G}$ and calibrated detuning for $V_\text{BG}=0$~V.}
		  \label{Fig1}
 \end{figure}

The device studied here is a fully-depleted silicon-on-insulator (SOI) nanowire transistor fabricated under CMOS standards. It consists of a 11~nm thick and 80~nm wide undoped Si (001) channel gated by a 50~nm long polycrystalline wrap-around silicon top-gate (G), as can be seen in Figure~\ref{Fig1}a,b. The SOI layer sits on a 145~nm thick SiO$_2$ buried oxide and a 850~$\mu$m handle wafer that can be used as a back gate~\cite{Roche2012}. The highly-doped source and drain are formed by ion-implantation, after deposition of 12~nm long Si$_3$N$_4$ spacers at both sides of the top-gate. A doping gradient occurs between the source-channel and drain-channel junctions producing confinement along the transport direction~\cite{Betz2014}. Furthermore, due to the corner effect in square cross-section nanowire transistors, accumulation happens first at the top-most corners generating a DQD in parallel~\cite{Sellier2007,Voisin2014,Betz2015,Gonzalez-Zalba2015}. 

In the presence of inter-dot tunnel coupling $\Delta_\text{c}$, the energy spectrum of a DQD with one electron presents a well-defined two-level system with an avoided crossing at zero-energy detuning ($\varepsilon=0$), as depicted in Figure~\ref{Fig1}c. At large detuning $\left|\varepsilon\right|>0$, the electron is strongly localized in one of the dots (left $\left|L\right\rangle$ or right $\left|R\right\rangle$ charge states). The ground (-) and excited state (+) energies of this system are given by

\begin{equation}
	E_\pm=\pm\frac{1}{2}\sqrt{\varepsilon^2+\Delta_\text{c}^2}.
\end{equation}
This two-level system has an associated differential capacitance, as seen from the top gate, given by

\begin{equation}\label{diff}
	C_\text{diff}=C_\text{geom}+(e\alpha)^2\frac{\partial\left\langle n\right\rangle}{\partial\varepsilon},
\end{equation}
where $C_\text{geom}$ corresponds to the DQD's geometrical capacitance and $\alpha$ is the difference between the right and left dot-to-gate couplings (see Supporting Information). The average electron occupation (defined here for the right dot) can be conveniently expressed as a function of the difference between ground state and excited state occupation probabilities $Z=P_{-}-P_{+}$ as

\begin{equation}\label{n}
	\left\langle n\right\rangle=\frac{1}{2}\left(1+\frac{\varepsilon}{\Delta E}Z\right),
\end{equation}
where $\Delta E= E_+-E_-$~\cite{Shevchenko2015}. Finally, using Eqs.(2)-(3) we arrive at the generalized expressions for the differential capacitance of a DQD,

\begin{eqnarray}
	C_\text{diff}&=&C_\text{geom}+C_\text{Q}(\varepsilon)+C_\text{T}(\varepsilon), \label{diff2} \\
  C_\text{Q}&=&\frac{(e\alpha)^2}{2}\frac{(\Delta_\text{c})^2}{\Delta E^3}Z, \label{quantum}\\
	C_{T}&=&\frac{(e\alpha)^2}{2}\frac{\varepsilon}{\Delta E}\frac{\partial Z}{\partial\varepsilon}. \label{tunneling} 
\end{eqnarray}

Expression~(\ref{diff2}) contains two contributions parametric on $\varepsilon$. The first, $C_\text{Q}$, corresponds to the so-called quantum capacitance arising from adiabatic charge transitions and the non-zero curvature of the energy bands~\cite{Sillanpaa2005,Duty2005}. The second, the tunnelling capacitance $C_\text{T}$, appears when population redistribution processes, such as relaxation and thermal excitation, occur at a rate comparable or faster than the probing frequency~\cite{Ashoori1992, Ciccarelli2012, Gonzalez-Zalba2015}. In general, both contributions must be considered when analyzing the effect of the qubit on an external system.

To detect the differential capacitance of the DQD, we use gate-based radio-frequency reflectometry~\cite{Colless2013,Betz2015,Gonzalez-Zalba2015,Urdampilleta2015} at the base temperature of a dilution refrigerator. We couple the DQD via the gate to a $f_\text{rf}=355$~MHz resonant tank circuit formed by a surface mount inductor ($L=390$~nH) and the gate to ground parasitic capacitance ($C_p=515$~fF). Additionally, a surface mount bias-tee allows us to apply a DC gate voltage ($V_\text{G}$). We apply a -95~dBm signal at the resonant frequency and monitor the phase of the reflected signal obtained from IQ-demodulation after cryogenic and room temperature amplification. The demodulated phase response is sensitive to capacitance changes $\Delta C$ of the probed system~\cite{Duty2005}, $\Delta\Phi\approx- 2Q\Delta C/C_p$ where $Q$ is the quality factor of the resonator ($Q=42$).

Figure~\ref{Fig1}d shows the demodulated phase response of the resonator ($\Delta\Phi$) as a function of the top-gate voltage ($V_\text{G}$) and back-gate voltage ($V_\text{BG}$) in the sub-threshold regime of the transistor, where direct source-drain current measurements are not sensitive enough (see Supporting Information). Here, we observe a diagonal line of enhanced phase response, which we identify with a single valence electron shared between quantum dots as demonstrated below. In this voltage regime, the tunnel rate between the source and drain reservoirs and the two quantum dots is slow, leading to a negligible reservoir-to-dot signals and indicating that both dots are well-centred in the channel~\cite{Petersson2010,Lambert2014b, Gonzalez-Zalba2015}. On the contrary, the inter-dot charge transition is still visible due to the finite tunnel coupling $\Delta_\text{c}$ between quantum dots and also due to a slight asymmetry in the dot-to-gate couplings, which could be due to potential irregularities at the interface~\cite{Voisin2014}. The inter-dot line is the last transition we observe, however excited-state spectroscopy revealed it is not the (0,1)-(1,0) transition but an odd-parity charge transition~\cite{Schroer2012} with total electron number higher than 1. Overall, these measurements demonstrate that gate-based reflectometry simplifies the qubit architecture and presents the advantage that charge motion can be detected without the need of direct transport or external electrometers.

In order to confirm the quantum nature of the inter-dot transition, we do a line-shape analysis of the signal, as can be seen in Figure~\ref{Fig1}e. Here, we plot the phase response as a function of gate voltage for $V_\text{BG}=0$. We use equation~(\ref{diff2}) to fit the data assuming adiabatic conditions for the inter-dot transition ($\Delta_\text{c}\gg k_BT_e, hf_\text{rf}$) since the electron temperature in the leads is $T_e<200$~mK. Under these conditions $Z\approx$ 1 and the differential capacitance of the system becomes only dependent of the tunnel coupling. We obtain $\Delta_\text{c}=98\pm 2~\mu$eV. Here, we have used $\varepsilon=e\alpha (V_\text{G}-V_\text{G0})$, where $\alpha=0.25$, accurately obtained from microwave spectroscopy measurements, as shown below, and $V_\text{G0}$ is the gate voltage value at which the signal is maximum.

\begin{figure}[htbp]
	\includegraphics{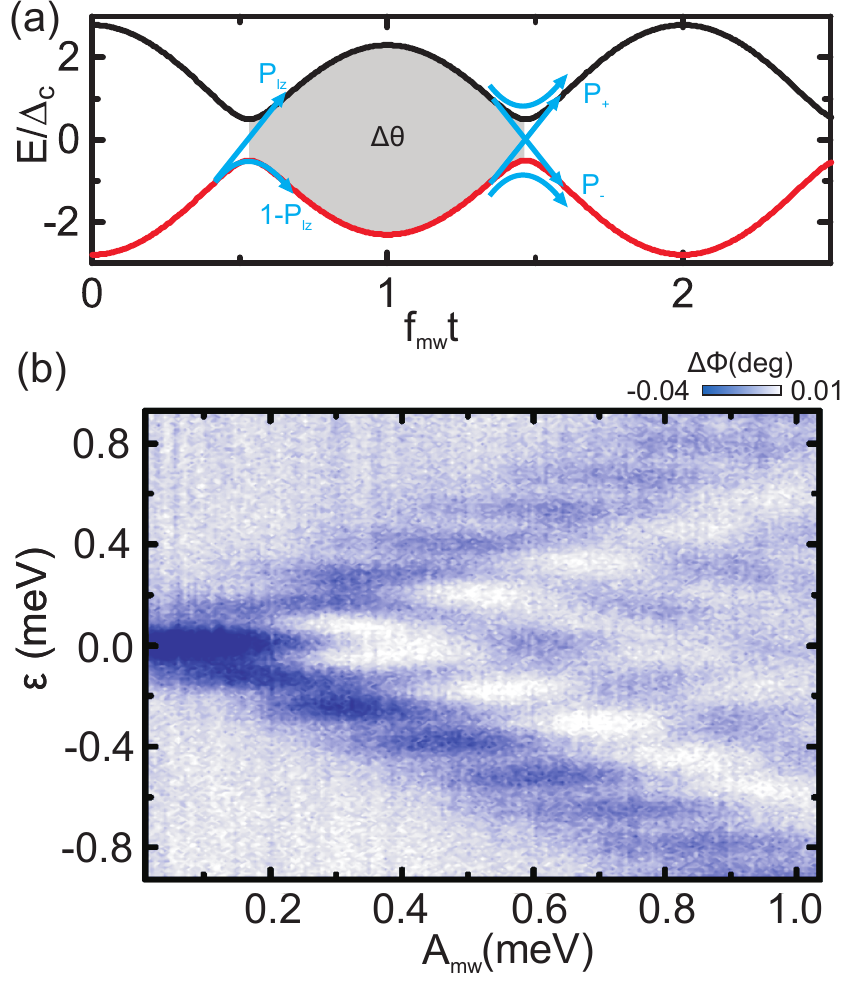}
	\caption{{\bfseries Dispersive detection of Landau-Zener-St{\"u}ckelberg-Majorana interference}. (a) Evolution of the ground (red) and excited (black) state energies as a function of time $t$, when periodically driven through the anti-crossing. Here Landau-Zener transitions occur with probability $P_\text{LZ}$. Between two successive transitions, the excited and ground states acquire a dynamical phase $\Delta\theta$ given by the area between energies. (b) Phase response of the resonator $\Delta\Phi$ as a function of detuning $\varepsilon$ and calibrated microwave amplitude $A_\text{mw}$.}
		  \label{Fig2}
 \end{figure}

We now move on to the investigation of microwave-driven coherent charge oscillations between quantum dots. Coherent transitions between the two charge states can be promoted by fast-oscillating voltage signals that vary the energy splitting periodically, as sketched in Figure~\ref{Fig2}a, where we plot the ground (red) and excited (black) state energies as a function of time. At the point of minimum energy splitting, a Landau-Zener transition occurs that splits the electron wavefunction in to ground and excited states with certain probability $P_\text{LZ}=exp(-\pi\Delta_\text{c}^2/2A_\text{mw}hf_\text{mw})$, where $A_\text{mw}$ and $f_\text{mw}$ are the amplitude and frequency of the driving signal respectively and $h$ is Planck's constant. After the first passage, the two states acquire a dynamical phase difference ($\Delta\theta$) given by the time integral of $\Delta E$, marked in grey in Figure~\ref{Fig2}a. If a second passage is performed at time-scales faster than the electron phase coherence time ($T_2$), a second Landau-Zener transition generates a quantum mechanical interference of the ground and excited state occupation probabilities $P_{-(+)}$. This phenomenon is known as Landau-Zener-St{\"u}ckelberg-Majorana interferometry, analogous to Mach-Zehnder interferometry~\cite{Sillanpaa2006,Oliver2005}, and allows probing coherent charge tunnelling and the time-scale at which they occur.

\begin{figure}[htbp]
	\includegraphics{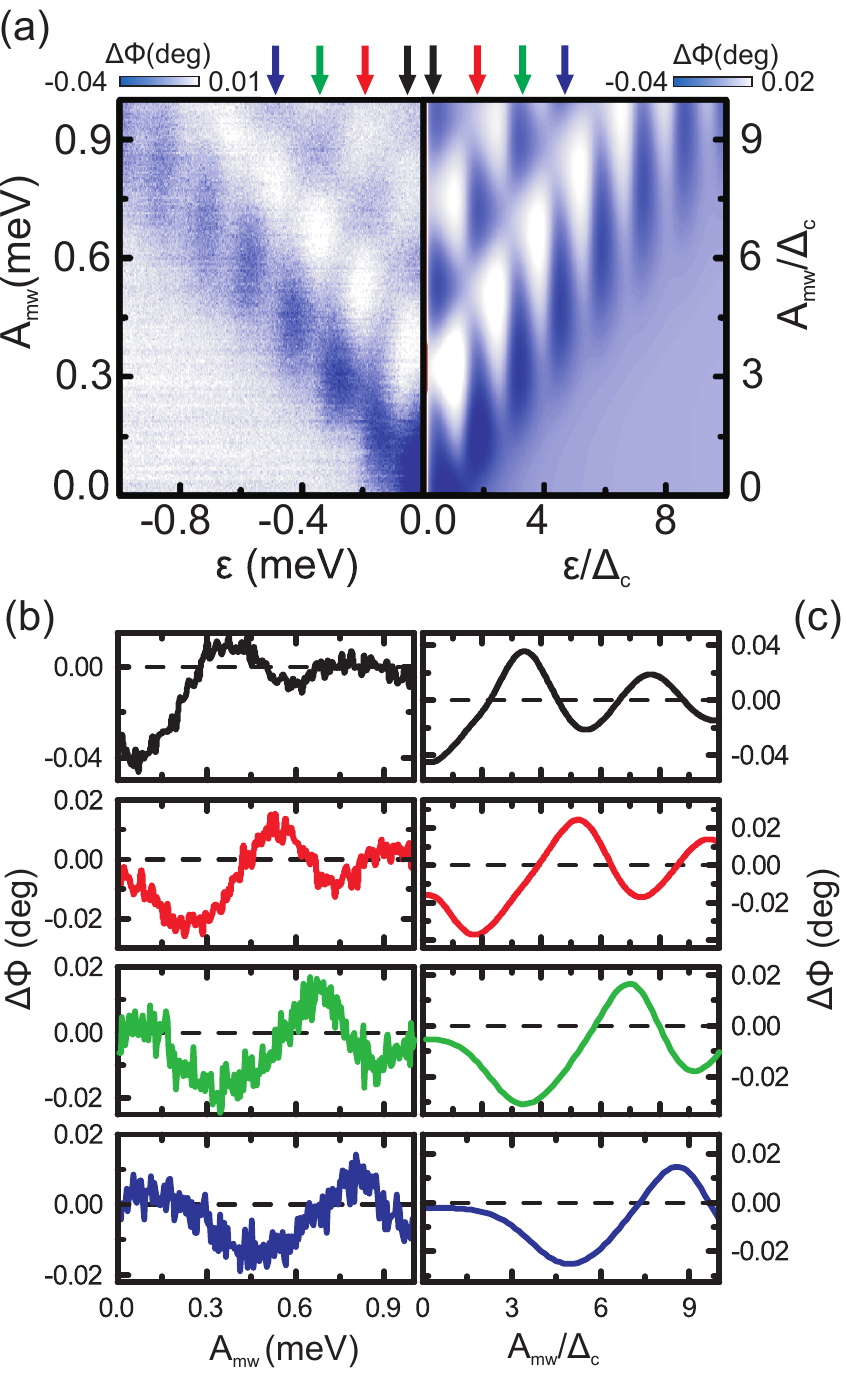}
	\caption{{\bfseries Theoretical analysis.} (a) Comparison between experimental and calculated LZMS interferograms. Experimental (b) and calculated (c) $n$-photon traces as a function of microwave amplitude. The traces are taken at the points indicated by the colored arrows in (a). The 0, 1, 2, 3 photon traces correspond to the black, red, green and blue solid lines, respectively.}
		  \label{Fig3}
 \end{figure}

To generate the required conditions to observe this phenomenon in our system, we vary the energy detuning periodically, $\varepsilon+A_\text{mw}cos(2\pi f_\text{mw}t)$, by applying an attenuated MW signal via the source of the device (see Figure~\ref{Fig1}a). Here we use $f_\text{mw}=34$~GHz and variable amplitude $A_\text{mw}=\kappa V_\text{mw}$, where we use $\kappa=0.46$~meV/V to calibrate the microwave generator output, $V_\text{mw}$ (See Supporting Information). The characteristic LZSM interference pattern is shown in Figure~\ref{Fig2}b, where we plot $\Delta\Phi$ as a function of detuning and microwave amplitude. In the region defined by $A_\text{mw}\geq\varepsilon$, the qubit is periodically driven through the avoided crossing which in turns affects the resonator response. First, we observe that $\Delta\Phi$ varies periodically as a function of $\varepsilon$, with resonant lines appearing at equally-spaced points $\varepsilon=nhf_\text{mw}$. Here, $n$-photon transitions mediate the charge oscillation between quantum dots and allow calibrating the dot-to-resonator coupling $\alpha$. Moreover, we see that, at fixed detuning, $\Delta\Phi$ oscillates (quasi)periodically around zero as a function of the microwave amplitude (seen in more detail in Figure~\ref{Fig3}b). Since $\Delta\theta$ is an increasing function of $A_\text{mw}$, what we observe here is the alternation between constructive and destructive interference in the ground state occupation probability. Overall, the results in Figure~\ref{Fig2}b demonstrate the dispersive readout of coherent charge oscillations in a semiconductor DQD via its interaction with an electrical resonator.  

Noteworthy are the regions of positive resonator phase shift in Figure~\ref{Fig2}b. In the simple adiabatic picture, the differential capacitance of a DQD simplifies to its quantum capacitance $C_\text{Q}$. Considering this limit, $\Delta\Phi>0$ implies an average population inversion which is not achievable in two-level systems. Understanding the qubit-resonator interaction in non-adiabatic regimes, such as LZSM, requires studying a hybrid regime in which not only quantum capacitance changes occur but also tunnelling capacitance variations.

We consider here the qubit-resonator system semi-classically: a quantum system coupled to a classical resonator $(hf_\text{rf}\ll k_BT$). Such a semi-classical approach was successful for the description of most phenomena related to atom-light interaction~\cite{Delone1985}. In our case, this assumption means that all characteristic qubit times are much shorter than the resonator period $f_\text{rf}^{-1}\gg h/\Delta_\text{c}, T_{1,2}$. Since the resonator is much slower than the qubit ($f_\text{rf}\ll f_\text{mw}$), it sees the stationary value for the occupation probabilities. Assuming this, we can make use of the analytic result for the time-averaged upper-level occupation probability $P_{+}$ in the strong-driving regime, obtained in the rotating-wave approximation~\cite{Shevchenko2010}

\begin{equation}\label{LZS}
	P_+=\frac{1}{2}\sum_n\frac{\Delta_{\text{c},n}^2}{\Delta_{\text{c},n}^2+\frac{T_2}{T_1}\left(\left|\varepsilon\right|-nhf_\text{mw}\right)^2+\frac{h^2}{4\pi^2T_1T_2}},
\end{equation}
where $\Delta_{\text{c},n}=\Delta_\text{c}J_n(A_\text{mw}/hf_\text{mw})$, and $J_n$ is the n$^{th}$ order Bessel function.

The differential capacitance of the DQD can then be calculated using equations~(\ref{diff2}),(7). We note that equation~(7) describes the series of Lorentzian-shaped multi-photon resonances, while its derivative gives the alternation of positive and negative values~\cite{Shevchenko2008}. Figure~\ref{Fig3}a shows the comparison of the measured (left) and calculated (right) LZSM interferometry patterns. Here, we use $T_2=100$~ps, obtained from Fourier analysis, as demonstrated below, and we use $T_1$ as a fitting parameter. We find the best fit for $T_1\approx T_2=100$~ps. These results justify our assumption that all characteristic qubit's times are much shorter than the resonator period. This short $T_1$ could be due to the presence of low-lying orbital excited states in the silicon quantum dots which have been reported to have relaxation times ranging down to the ps regime~\cite{Tahan2014}. 

To further demonstrate the good match between experiment and theory, we plot, as a function of the amplitude of the MW signal, the measured and calculated $n$-photon traces in Figure~\ref{Fig3}b,c respectively. The $n=0,1,2,3$ (black, red, green, blue) are obtained at the points marked by the arrows in panel (a).  We observe that the theory successfully captures the oscillatory behaviour of the differential capacitance, highlighting the importance of the third term in equation~(\ref{diff2}) and sets LZSM in a regime where quantum and tunnelling capacitance changes must be considered. The match is particularly good for the $n=1,2,3$ photon lines while for the 0-photon line the agreement is qualitative. This can be understood knowing that equation~(7) assumes $\Delta \ll \left\vert \varepsilon \right\vert $, which means that it is not exact at around $\varepsilon =0$. Nevertheless, its practical implementations~\cite{Dupont-Ferrier2013, Forster2014, Silveri2015} demonstrated that this gives reasonable description even for $\varepsilon \sim \Delta $.

\begin{figure}[htbp]
	\includegraphics{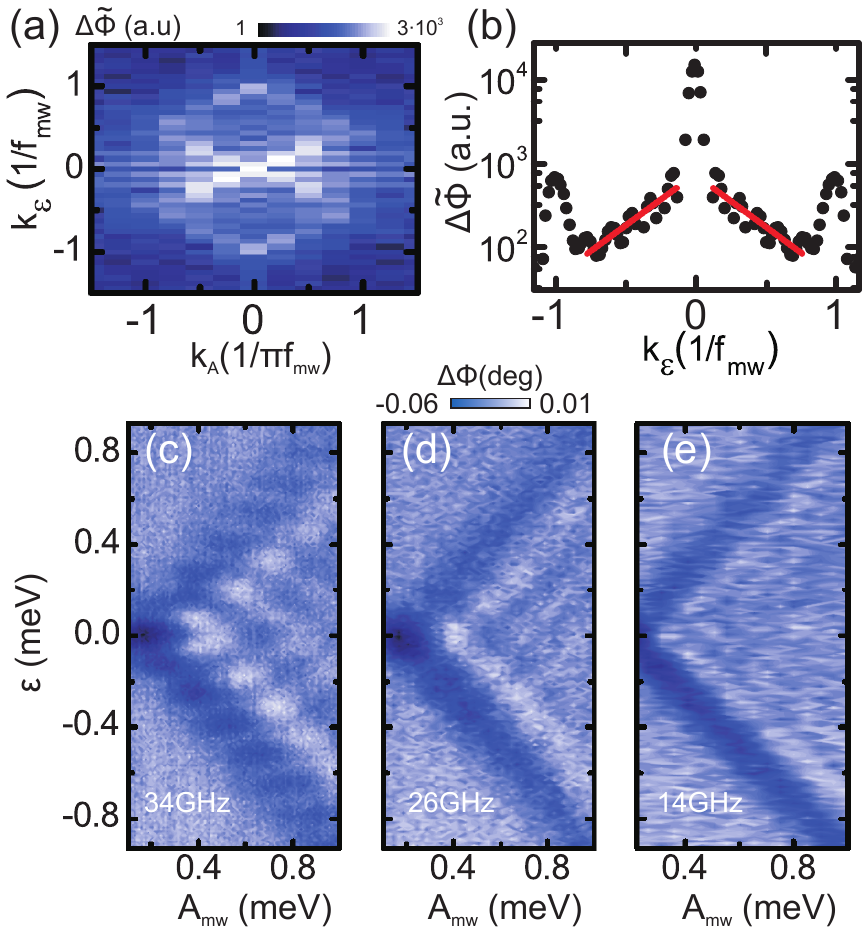}
	\caption{{\bfseries Coherence Time}. (a) 2D numerical Fourier transform of Figure~\ref{Fig2}(b) centred around the first reciprocal zone. We plot the transformed phase response $\Delta\tilde{\Phi}$ versus the reciprocal-space variables $k_\varepsilon$ and $k_\text{A}$. (b) $k_\text{A}=0$ trace showing the exponential decay of the transformed phase response. LZSM interference patterns taken at different microwave frequencies $f_\text{mw}$: (c) 34~GHz, (d) 26~GHz and (d) 14~GHz.}
		  \label{Fig4}
 \end{figure}

Finally, we move on to the study of electron phase coherence time in our strongly-driven two-level system. In Figure~\ref{Fig4}a, we perform a Fourier analysis of the dispersive response of the resonator of Figure~\ref{Fig2}b. The two-dimensional Fourier transform of the phase response, $\Delta\widetilde{\Phi}$, shows the characteristic lemon-shaped ovals of increased intensity in the reciprocal space ($k_{\varepsilon}, k_\text{A}$) similar to results obtained for superconducting qubits~\cite{Berns2008} and semiconductor quantum dots~\cite{Cao2013,Forster2014}. Two-dimensional Fourier transforms of the occupation probabilities in the LZSM regime have been demonstrated to carry information about the qubit's dephasing mechanisms. More particularly, the transformed populations decay exponentially in $k_\varepsilon$ as exp$(-k_\varepsilon/T_2)$~\cite{Rudner2008,Cao2013}. This result is directly applicable to $\Delta\widetilde{\Phi}$ since its associated differential capacitance is proportional to the occupation probabilities, $P_+$ and $P_-$ through the quantum capacitance term as seen in equation~(\ref{quantum}). We demonstrate this in Figure~\ref{Fig4}b where a one-dimensional $k_\varepsilon$ trace at $k_A$=0 reveals an exponentially attenuated signal. From the fit, we find $T_2=100\pm$50~ps, similar to values reported for charge coherence in semiconductor double-dots~\cite{Dupont-Ferrier2013, Shi2013} and Cooper-pair transistors~\cite{Nakamura2002}. 

We confirm our estimation of $T_2$ by performing a frequency dependence of the LZSM pattern in Figure~\ref{Fig4}c,d,e. Here we explore three driving regimes: the quantum coherent regime [panel (c)] the incoherent driving regime [(e)] and an intermediate driving regime [(d)]. In the quantum coherent regime, measured at $f_\text{mw}=34$~GHz, successive transitions through the anti-crossing are correlated and we observe the clear signature of the interference fringes indicating $f_\text{mw}>T_2^{-1}$. In the incoherent regime, $f_\text{mw}=14$~GHz, Landau-Zener transitions are uncorrelated and we observe no sign of interference oscillations, hence $f_\text{mw}<T_2^{-1}$. However, in the intermediate regime, $f_\text{mw}=26$~GHz, we observe only one clear minima and maxima regions, indicating that the number of correlated passages is close to two and hence $f_\text{mw}\approx T_2^{-1}$. These results agree well with the coherence time obtained from the Fourier analysis.

In conclusion, we have reported the dispersive read-out and coherent manipulation of a DQD in the channel of a CMOS nanowire transistor. Gate-sensing allows for \textit{in-situ} detection of charge motion within the double-dot system without the need of external electrometers. Additionally, we have performed coherent manipulation of the DQD charge state by means of high-frequency microwave signals and observed the emergence of LZSM interference in the resonator's response. Furthermore, we have demonstrated that, in fast relaxing systems, the dispersive DQD-resonator interaction contains contributions from both the quantum capacitance and the tunnelling capacitance.
In the future, split-gate CMOS transistors, as the ones reported in refs.~\cite{Dupont-Ferrier2013,Betz2015}, could provide better control of the energy detuning between dots and a larger asymmetry in the dot-resonator coupling, improving the sensitivity of the read-out protocol. Overall, our results demonstrate that it is possible to integrate qubit control and read-out with existing CMOS technology opening a path towards large-scale integrated qubit architectures.


\begin{acknowledgement}
The samples presented in this work were designed and fabricated by the TOLOP project partners, http://www.tolop.eu. This research is supported by the European Community's Seventh Framework Programme (FP7/2007-2013) through grant agreement No. 318397. This work is partially supported by the RIKEN iTHES Project, the MURI Center for Dynamic Magneto-Optics via the AFOSR award number FA9550-14-1-0040, the IMPACT program of JST, and a Grant-in-Aid for Scientific Research (A). A.J.F. was supported by a Hitachi Research Fellowship and acknowledges funding from EPSRC under the grant EP/K027018/1.
\end{acknowledgement}

\begin{suppinfo}
Detailed equations of the differential capacitance of a double quantum dot, additional DC transport measurements, calibration of the microwave amplitude and excited-state occupation probabilities in the LZS regime.
\end{suppinfo}

Note: The authors declare no competing financial interest.

\bibliographystyle{achemso} 

\end{document}